# The influence of Automated Decision-Making systems in the context of street-level bureaucrats' practices

Author 1: Manuel Portela[1]*

Author 2: A. Paula Rodriguez Müller [2]

Author 3: Luca Tangi [3]

[1] Universitat Pompeu Fabra, Barcelona, Spain

[2] Joint Research Center, European Commission, Seville, Spain

[3] Joint Research Center, European Commission, Ispra, Italy

*Corresponding author: manuel.portela@upf.edu

Abstract

In an era of digital governance, the use of automation for individual and cooperative work is increasing in public administrations (Tangi et al., 2022). Despite the promises of efficiency and cost reduction, automation could bring new challenges to the governance schemes. Regional, national, and local governments are taking measures to regulate and measure the impact of automated decision-making systems (ADMS). This research focuses on the use and adoption of ADMS in European public administrations to understand how these systems have been transforming the roles, tasks, and duties of street-level bureaucrats. We conducted a qualitative study in which we interviewed street-level bureaucrats from three administrations who had used an ADMS for several years, which was embedded in their daily work routines. The outcome of our research is an analysis of five dimensions of how collaborative work, the organizational settings, the capacities of bureaucrats and the implementation of the ADMS enable or limit the capacities for offering better services towards the citizens.

---

Policy Significance Statement

Our main conclusions are drawn from studying the challenges and opportunities of ADMS in the public administration looking at how these systems are adopted by their users and implemented in real-case scenarios. We present a taxonomy that could be used to assess the transformation of public sector with the introduction of ADMS and novel AI systems.

---

1. **Introduction**

The green and digital transition has been spreading at a fast pace during the last decade with particular development of solutions based on data processing and advanced software for automation to offer more efficient and cost-effective services. Automated decision making systems (ADMS) are algorithm-based systems that provide some level of automation for specific tasks, usually done by humans, to increase the efficiency or accuracy of the decision making process, such as triaging, scoring or pattern detection.

ADMS are attractive solutions for the public sector, offering a novel approach to making complex decisions that affect large populations. Some cases are the allocation of social benefits in the welfare system, risk assessment in the context of policing or criminal offence, or the offering of the correct assistive path to help jobseekers overcome their unemployment situation.

Despite the increasing presence of ADMS in public administration, there are several challenges around human involvement in algorithmic systems, mainly related to trust and efficacy during its use and implementation (Portela and Alvarez 2022; Wagner 2019). **This research focuses on the adoption of ADMS by street-level bureaucrats to understand their perception of the systems and draw learnings from their developing capacities in the context of digital transformation.**

This article is organized as follows. First, we will address the context and the scope of the research. Second, we address the state-of-the-art in the context of ADMS in general, and its specific use in the public sector. Third, we delineate our objectives and contributions. Fourth, we explain the methodologies used in the research study. Fifth, we describe our findings. Last, we offer some conclusions and recommendations.

### 1.1. Scope and context

The notion of ADMS considers a broad implementation of multiple decision points, data sets, and technologies used to automate a decision-making process in a range of contexts, including public administration, business, health, education, law, employment, transport, media and entertainment, with varying degrees of human oversight or human intervention. In the context of the European Union, the first provision about the use of ADMS is in Article 22 of the General Data Protection Regulation (GDPR).

In the scope of this research, we approach ADMS as a general approach to automating processes. As Suski (2021) proposes, ADMS could be expert systems, rule-based or machine-learning software. These are different branches of AI in the computer science field but are less known in common understanding of what AI is. In current debates, AI is associated with Machine Learning, or more specifically, deep learning based systems. These are different AI applications, such as those meant to merely provide information (e.g. chatbots), provide biometrical identification or govern systems to process information in an automated manner (automatic reasoning). Rather, ADMS are commonly simple algorithm-based systems that perform tasks of triaging, classification, risk assessment, or data summarization used to facilitate, inform or automate decision-making processes. To avoid such confusion, we approach the terms from the perspective of the purpose of an ADMS, which is automating a process. For this reason, we will use the term AI indistinctly and consider literature that may refer to AI or ADMS focusing on the purpose of the system.

The report "European landscape on the use of artificial intelligence by the public sector" (European Commission. Joint Research Centre. 2022) collects and summarizes the AI national strategies to date in European countries. These strategies are meant to stimulate awareness, improve data access and quality, improve internal capacity, collect lessons learnt from pilots, include ethical and legal AI guidelines and, specify funding and procurement procedures. The increase of AI and ADMS in the public sector is evident. In 2022, it was reported that from the total AI implementations, 38% of AI solutions were already implemented, while 25% were in the pilot stage and 30% were pilots, proof of concepts or experimentations. The report acknowledges 58% of implementations of AI that are merely Machine Learning algorithms. Other relevant uses include automated reasoning, planning and scheduling or natural language processing.

### 1.2. Use of ADMS in the public Administration

ADMS have been recently studied in the public sector under perspective on the law and management fields (Schäferling 2023; Suksi and Suksi 2023). An ADMS could bring full or partial automation to a process and is commonly placed in-between non-automated processes (Birhane 2021; Cummings 2004, 2006; Jones 2017; Wagner 2019; Zarsky 2016).

Following the EC's report, uses of AI in public administrations cover issues of Public order and safety purposes, service personalization services (e.g. for Health or Social protection), prediction and planning in economic affairs or information analysis and prediction and planning in education, environment and cultural or recreation sectors. Other studies acknowledge how administrations have already started adopting AI as ADMS (Ahn and Chen 2020; Misuraca and Noordt 2020; Wirtz *et al.* 2019) in several different areas such as policing and surveillance, and other service delivery (Zuiderwijk *et al.* 2021).

Besides the potential benefits, the use of AI and ADMS, they pose several risks that should be acknowledged (Zuiderwijk *et al.* 2021). For example, data practices in several ADMS resulted in a disproportionate way harming vulnerable social groups (Biber 2023; Ruijer *et al.* 2023). Besides, biased data can also affect the outcome of the algorithms and the decision process (Alon-Barkat and Busuioc 2023; Mosier *et al.* 1998). Data used by algorithms can be critical and controversial, despite the good intentions of providing more social equity.

Previous studies acknowledged that the implementation of ADMS is not equal among the countries and it requires changes in the organizational structures (Weerakkody *et al.* 2011). However, adopting such systems is not always motivated by the size of the offices. Instead, because of its complexity, it requires expertise in digitalization and to tackle sector-specific obstacles (Neumann *et al.* 2024). Other challenges in its implementation include the lack of safety and robustness of systems, financial feasibility, or the need to integrate several already-existing systems (Wirtz *et al.* 2019).

Beyond these identified opportunities and challenges in the development and implementation of ADMS, several barriers occur when these systems are put in place and during its use. In this article, we focus on the ADMS in practice, to learn how organizational and technological settings affect the use of ADMS and what street-level bureaucrats can tell from their experience.

### 1.3. ADMS in practice

ADMS systems are more than an automation mechanism. Instead, we should consider them as socio-technologies because they "have an impact on people, inter-personal interactions, and society as a whole, as they are able to recognize these social components of their environment" (Biber 2023, 13). Therefore, the design, development and implementation of such systems should not be separated from the situation of use.

As ADMS are meant to be placed in the context of decision making, these systems may involve direct interaction. Article 22 of the GDPR prohibits the implementation of fully-automated decision making (Biber 2023; Binns and Veale 2021), leaving only space for partial automation. In this context, the capacities and skills of civil servants are key to interpreting and adopting ADMS outputs (Green and Chen 2020). The role of human operators in decision-making can be critical in particular scenarios, such as border control (Noori 2021), social welfare (Ranerup and Henriksen 2022) or criminal justice (McCallum *et al.* 2017), where the decision could limit the rights and benefits for an individual or an entire social group.

Depending on the type of ADMS, humans can play different roles in the decision-making and various forms of using algorithms and machine mediation (Binns and Veale 2021). However, for studying such roles, it implies a broad understanding of how humans behave and interact with machines along with an understanding of the cultural and societal constraints and biases that algorithms reproduce. For humans to act accordingly, it is also needed to consider other human and contextual factors such as system transparency (Ananny and Crawford 2018; Kemper and Kolkman 2019), individual confidence and motivation (Zhang *et al.* 2020) and trust in the automated system (Lee and See 2004), among others.

Regarding the operational environment, human decision-making is impacted by the interaction with the technological artefacts and their interfaces. Algorithms are fed with historical data that could be structurally biased, and methods used for training the algorithms tend to reproduce such biases (Barocas 2014). But humans also can have their individual biased preferences that affect decision-making (Goodman 2016) or follow system recommendations only if confirm what they already thought (Selten *et al.* 2023). In the decision moment, humans interact with algorithms through a machine and its interface but are influenced by several environmental conditions and human factors such as workload, personal beliefs, motivations, capabilities, knowledge, experience, comprehension and attention abilities (Cranor 2008; Kulju *et al.* 2019; Noori 2021).

An additional barrier to adopting ADMS is related to the perception of the system and its outcome. Trust, reliance and confidence are different social constructs that can affect the relationship between the operator and the system (Vereschak *et al.* 2021). It can be tempting to think that the accuracy of a system might be key to creating confidence in the system. However, this is not always the case as confidence can vary depending on their individual experience (Portela *et al.* 2024). Moreover, a recent study found that narratives influence expectations, making users feel more confident, take more risks and perceive a superior performance working together with the system (Kloft *et al.* 2024). But this overreliance on the system can lead to what is called "rubber-stamping" and losing control and oversight of the systems' errors and potential damages (Wagner 2019).

The perception of the reliability of a system is tight to the experience of use and its context, as demonstrated in a study done after a scandal about the use of an unfair ADMS (Alon-Barkat and Busuioc 2023). But a more transparent or explainable system does not necessarily become more trusted by humans (Portela and Alvarez 2022), it needs to be more accessible (Grimmelikhuijsen 2023). Accessibility is the availability of trusted information about the data, the code and the models used to train the algorithm.

For human operators to be able to give a discrete and justified response, they might have a different level of expertise in the context of use, from long-experienced professionals to operators without expertise (Myers-West *et al.* 2019). Users can be able or not to use an ADMS with more or less success depending on how the system provides relevant information and their expertise in understanding data and visual communication (Rudin 2019; Wang *et al.* 2020).

But, sometimes public administrations may fail to provide sufficient training and tools to decision-makers and civil servants needed to develop new skills and capacities to use the ADMS (Wihlborg *et al.* 2016). In consequence, affecting the perception of civil servants towards the system, decreasing their trust in AI systems and their own confidence with a high impact on the public opinion (Fergusson 2014).

Our analysis of previous works acknowledges several challenges that can influence the use of an algorithm-based system in automated settings. These include the operational context, the human capacities, the perception of the system, the transparency of the system and the confidence towards the system's output. Consequently, the scope of this research is to address these challenges and study how capacities of civil servants and street level bureaucrats could be increased to offer better services with the adoption of ADMS.

## 2. Research questions

In light of the above, this work addressed two research questions:

**RQ1:** What skills and competencies do street level bureaucrats and public servants need to perform their tasks while supported by the ADMS?

**RQ2:** What are the different types of technological, social, and organisational arrangements that facilitate or hinder the use, implementation and operation of ADMS in public administrations?

The main expected outcomes related to mapping the different ways ADMS are used in context result from the following:

- Build a narrative about how ADMS are perceived by their users;
- Classify types of approaches, weaknesses and strengths in algorithmic systems; and
- Identify outcomes of using ADMS in public administrations and distil best practices.

## 3. Methodologies

Our research engages theoretically with *Infrastructuring Studies* as an interdisciplinary approach crossing the fields of Information Systems, Science and Technology Studies and Computer Supported Collaborative Work. We take the theoretical framework of *infrastructuring* "as a lens to understand how processes of reality-making actually take place" (Parmiggiani 2019). There are certain key infrastructural aspects of information systems: Interconnectedness and complexity, Layer approach and standardization; and, (In-)Visibility in use. By understanding these, "highlight aspects of design methodologies that have less to do with designers/developers and their design process, and more to do with how the technologies undergoing design, and the design process itself, are embedded in an existing work environment." (Pipek and Volker 2009, 449).

Following the authors, there are many approaches to frame and study infrastructures. As a form of collecting evidence for our *infrastructuring* approach, we make use of a particular branch of participatory design which is the study of artifact ecologies (Bødker *et al.* 2017; Jung *et al.* 2008). Based on the assumption that artifacts cannot be fully understood individually, artifact ecologies empirically explore the relationships between interactive artifacts in people's personal life. This approach has been conceived as a theoretical framework "to address the gap between culture, experience and the practical role of artifacts in embodiment and mediation" (Bødker and Klokmose 2011, 316). Artifacts ecologies look at goal-oriented action and how artifacts mediate and enact endless possible actions of the environment.

As Maragno et. al. (2023) argue, affordances and constrains in the use of AI in the public sector are less explored. The artifact ecologies not only analyses the interaction with an artifact (i.e. a computer interface) but the entire ecology of artifacts and their environment, and the dynamics that are created in-the-making while administrative staff develop their work.

### 3.1. Tools and methods

For this research, we selected three cases in Europe that have been implementing an ADMS for at least two years. The cases were selected from the Public Sector Tech Watch (PTSW) database[1], that has more than 1,200 cases on the use of AI in the EU public sector. We preselected a list of 20 cases, and interviewed different administration managers until we found those that fit with our requirements. In each case, we asked public administrations to give access to their staff and street level bureaucrats to be interviewed. The study design is based on field-oriented research with qualitative data collected through interviews, ethnographic observations and document analysis (Garton and Copland 2010; Rapley 2001; Silverman 2017).

The research was developed in two stages. First, individual interviews to street level bureaucrats with assistive tools, such as Likert-scales and mind maps to assert their perception of the system, their confidence and devices that are part of the decision making process (Bødker and Klokmose 2011, 2012). Second, we carried out a validation focus group with the participants to discuss the previous findings and expand the results using participatory methods for understanding how infrastructures support the field of work in an organization (Pipek and Volker 2009).

#### 3.1.1. Initial Interview

As an exploratory practice, a first interview was conducted with a representative from the use-case administration. The goal of these interviews was to collect details about the case and to communicate the purpose of the research in order to find the right people within the organization for the next stage in the study. A preparation guide was created for this purpose.

#### 3.1.2. Semi-Structured Interviews

For each case, we target to interview at least three street-level bureaucrats. The goal of these interviews was to learn deeper about the use of the ADMS and the environment in which these systems are inserted. Administrative staff can be very informative on the dynamics, frictions and behaviours drawn upon the experience of using a system.

Semi-structured qualitative studies consist on involving interviews and observations, with some explicit structure in terms of theory or method, but are not completely structured. In this set of studies, there are no previous hypotheses but themes emerge from the data. This implies that the involvement of the researcher and the interviewees plays a fundamental role in the results of the study. Semi-structured interviews are a key method for understanding people's perceptions and experiences (Blandford 2013).

Interviews took place in December 2023 (VioGen), February 2024 (NAV) and April 2024 (Slimme Check). After our interviews, we categorized and classified the qualitative information through a thematic analysis. The analysis dropped a series of common topics grouped in these five dimensions: Capacities, Organization, Implementation, Use and Confidence.

#### 3.1.3. Focus groups

After the information from the interviews was processed, a validation focus group was carried out. The goal was to validate the findings during the individual interviews and to put in common the main concepts. This process was also important to let the staff exchange their ideas and elaborate more on their own thoughts and beliefs about their individual experience. The main outcome was to validate the conceptual framework that served as the theoretical discussion in the overall research project.

The focus group was prepared to contrast the findings with the participants in case they wanted to expand or clarify any of the information. The focus group lasted around 2 hours, and included a presentation with results from the interviews and a space for interaction where participants discussed their perception of the findings and the typologies.

---

[1] European Commission, Joint Research Centre (JRC) (2023): Public Sector Tech Watch latest dataset of selected cases. European Commission, Joint Research Centre (JRC) [Dataset] PID: http://data.europa.eu/89h/e8e7bddd-8510-4936-9fa6-7e1b399cbd92

The focus group was split into four sections:

- A general introduction of participants and the cases.
- The presentation of main findings from interviews where participants could comment on each of the findings with their views and opinions.
- A two-stage questionnaire where participants were invited to agree or disagree with different affirmations. After they answered, they commented on their answers. The first stage was about skills and competencies; and the second stage was about technological, social, and organizational arrangements.
- A short list of recommendations was presented, and participants were able to create their own and comment their views about them.

Participants from all three cases were invited to participate in a Doodle to set up their convenient date several weeks in advance. Individual invitations were sent to participants to attend the focus group.

## 4. Case Studies
### 4.1. VioGen (Spain)

VioGen[2] was developed under the Spanish organic law 1/2004 and consists of a web-based system that is accessed by users of the nation-wide forces and corps of security but also penitentiary institutions, courts, medical institutes, legal and forensic sciences, and many others. This application works as a platform to exchange information between its users, who can access different levels of information about the case. The system has two main components that allow professionals to assess the situation to make a decision: the police level assessment risk of violence against women (VPR) and the forensic level assessment risk of violence against women (VFR)(López-Ossorio *et al.* 2019).

The implementation of the system includes continuing training of different local, regional, and national forces and bodies in the use of the system. A manual is also included in the provision of this system. By 2023, more than 6 million cases were evaluated and more than 92 thousand are active cases. There are many benefits acknowledged. In the first stage, it promotes and facilitates coordinated actions between different institutions involved. It offers a unified database about domestic and gender violence, reaching all the public entities. The different forms help to unify in a single protocol and standardize the response along the national territory. Lastly, it offers a system of alarms that helps to keep track of cases with different levels of risk.

Challenges to its implementation are related to the continuous changes in the nature of the crimes. The definition of domestic and gender violence responds to the definition in the Spanish law 1/2004. Regarding the skills required for using the system, not all users are trained equally in the territory. It also depends on the number of available members in different locations. To protect personal data and the privacy of victims, as mandated by the data protection regulation, each type of user has differential access according to their duties and responsibilities, so they can only access a limited amount of information available in the system. Because VioGen is a support decision system, police staff can increase the level of risk by hand and new evidence if they consider it. However, the level of disagreement is less than 15% in the lower levels.

### 4.2. NAV algorithm (Norway)

The Norwegian Labor and Welfare Administration (NAV) provides a single gateway into public labour and welfare services. Unemployed people that have problems in finding a job will receive help from NAV. Nevertheless, helping the population can be resource demanding and can lead to a long waiting period.

The implemented rule-based algorithm[3] allows classifying jobseekers into different categories every time that a person requests assistance to find a job or requests unemployment benefits. Based on Section 14a of the NAV Act, NAV categorizes users according to how much help it is assumed the person needs.

---

[2] Accessed at: https://www.interior.gob.es/opencms/es/servicios-al-ciudadano/violencia-contra-la-mujer/sistema-viogen/

[3] Accessed at: https://arbeidogvelferd.nav.no/news/2022/6/Algoritmer%20ga%20arbeidsledige%20vedtak%20om%20mer%20hjelp%20fra%20NAV

The categories are four: standard, situational, especially adapted, and permanently adjusted. Where the former are the simpler cases, and the latter are more complex and have a reduced ability to work. The proposed categories are based on jobseekers' age, education, stated health challenges and recent work history.

In 2018, when profiling using algorithms was introduced, a third of the NAV supervisor's decisions coincided with the algorithm's advice on *situational* category, which is somewhat increased effort. A year later, the NAV supervisors followed the algorithm's advice on *situational* category in half of the cases.

In 2023, the national unit was established. It is a virtual unit, meaning all processes and interactions are carried out through digital means. It has approximated of 30 caseworkers who are located across Norway. Most of them used to work at a local NAV office before they started working at the national unit. The unit is set up to handle *standard* users, age 30-59 and who have sufficient digital skills. The unit receives cases from the algorithm and from the local offices, and the caseworkers assess if the assumed standard users are indeed standard users or should be transferred to the local office. They have a turnover of about 2500 users a month. According to caseworkers, it is more convenient and efficient to respond to these cases online that do not require special attention.

Challenges are various. The algorithm can be improved making it more robust regarding the decision. Although, the incorporation of new data to have a more complete algorithm is a difficult challenge due to the data protection regulation. Another challenge is to improve the interface and the tools available. Automating some tasks could improve caseworkers' job by automating these tools, since all communications and follow-ups are made manually. It might improve the efficiency of the case workers, but a more comprehensive training will be required.

Since all the caseworkers have long experience in the local offices and know how to identify issues in each of the jobseekers, it is unclear how this system can be as useful for new case workers without previous experience.

### 4.3. SlimmeCheck (Netherlands)

The municipality of Amsterdam provides social assistance benefits. However, not everyone who applies for social assistance benefits is eligible. That is why the municipality is investigating assistance applications that may be unlawful. Amsterdammers who are entitled to assistance are also sometimes examined by an employee of Enforcement Work and Income. The idea was to test whether an algorithm can help determine which applications should and should not be investigated. So, Amsterdammers may suffer less from the burden of the enforcement process and the provision of assistance remains affordable.

The SlimmeCheck (Smart Check) algorithm[4] supports the employee to determine whether a livelihood application is worthy of research. The algorithm makes transparent and explainable which data led to the label 'worth researching'. All data used to come to an assessment is documented and described.

The municipality run a pilot with a new working method from April to July 2023. In this pilot, the algorithm determines whether an application is *researchworthy*. This is then checked by an Enforcement's office employee. The algorithm can find connections and patterns in a large amount of information about the assistance benefit requested and determines which information can be associated more often with applications for which further examination was necessary and which was not. SlimmeCheck was trained on historical data and consists of fifteen data points trained under a XGBoost algorithm. The outcome of the system includes also the three main features that are critical to make the decision.

Results from the pilot implementation showed that on almost all sensitive characteristics (such as age, country of birth and nationality) the developed model treats different groups more equally compared with the manual procedure. The model can better estimate which applications are worthy of research than an employee. As a result, capacity can be better used, and fewer unlawful benefits are provided.

Because the model can better estimate which applications need to be checked extra, it is less common for applications to be unnecessarily examined additionally. This leads to better proportionality. It also prevent the unlawful provision of benefits more often, which leads to a more preventive rather than a repressive approach.

### 5. Typologies

---

[4] Accessed at: https://algoritmeregister.amsterdam.nl/onderzoekswaardigheid-slimme-check-levensonderhoud/

After our thematic analysis and validation, our findings are three-fold. A set of typologies have been identified together with specific roles and organizational structures, which constitute the effects of infrastructuring in the context of each use case. The goal of identifying typologies among street-level bureaucrats was to explain how they interact with the ADMS and how decisions are made.

### 5.1. Dimensions

The identification of typologies requires to research in depth how they use the systems in the context of use. Nevertheless, due to the limitations of this research where only a few interviews per case were made we only grasped the general characteristics of how these typologies work. As a result, we obtained five dimensions on how ADMS impact, and six roles that interact in the context of the systems.

**Capacities:** Knowledge and skills developed to use the system, their expertise, and the resources to make a decision. Several authors raised the importance of having the right capacities of making right decision and interpret data, specially risk levels (Lipkus *et al.* 2001; Scurich *et al.* 2012).

**Organization:** Organizational structures that coordinate and delegate tasks within the administration, participatory processes, and supervision. Organizational settings could empower or constrain human decision but also affects the development and implementation of ADMS along the organization.

**Implementation:** The transformation of the administrative process to make place to the system, including how data is collected and the output of the system. Designing and Implementing the ADMS in the context could foster the adoption or limit the action of the decision-making process influencing in the quality of the service.

**Use:** The contextual and situational settings defined above condition how the system is used. Because no system is implemented in isolation, collaboration with other people and other technologies that are part of the daily work should be considered also part of the system.

**Confidence:** The perception of the system could increase its use and how efficient the administration becomes. Confidence and trust are key social constructs to understand adoption of an ADMS. Thus, the level of confidence could facilitate the system's adoption and affects the decision process.

### 5.2. Roles in the context of ADMS

During the interviews we spoke with different participants that, despite being system users or not, had dissimilar roles. These roles are exchangeable, and a person could have overlapping roles:

**Project leader or product owner:** usually this is a directive and technical role that supervise and lead the implementation or the design of the system. It could be a past user of the system, but it is unexpected to become a daily user.

**Team leader:** This role coordinates different other roles and has the capacity to coordinate the workload imbalances generated by daily work. A team leader could implement measures to increase capacities, such as training, communication tools or request new features in the system. The team leader could also be a daily user and supervises those extreme or sensitive cases.

**Researcher:** This role could be covered by someone with several years of experience in the administration or using the system; it helps to improve or implement the system and has the knowledge on how the process work. This role has a close relationship with other users and perform supervision on sensitive or extreme cases.

**Decision-makers:** A decision-maker is the role that receives the output of the system and should make a decision. Some ADMS have an output (e.g. risk assessments like VioGen) and others do not (e.g. triage systems like Slimme-Check or the NAV algorithm); therefore not always a decision-maker is needed. This role requires being a domain expert and be trained on the use of the system. Most decision-makers should follow a common criteria. Special training is required to avoid over- or under-reliance towards the system.

**System user:** A system user in the context of the ADMS is the role that handles information, it could on its input or its output, but not necessarily to interpret or to make a decision-regarding the ADMS output. This role can handle this information and request a review on the system decision or more explanation.

**End-users:** Commonly a role occupied by citizens but is not always the case. End-users are beneficiaries of the decision and, therefore, systems should be designed to enrich their experience and being fair towards them. In the case of decision-makers become end-users, these interpret and mediate with citizens. In both cases, usability is crucial and depends on how well designed the system is to correctly collect the right data and provides an efficient solution.

### 5.3. Organizational structures

Following roles, we characterize different organizational features:

**Managerial:** A unit is defined to manage the development, research or implementation of the system, this unit can be centralized or decentralized at different levels.

**Deployment:** A system can be defined to be used by a particular group of bureaucrats or be distributed into an existing organizational structure.

**Territorialization:** bureaucrats can be centralized in a single territorial place, working remotely as a single virtual unit, or can be distributed in local offices

**System-user interaction:** The system can be accessed by one type of bureaucrats or by different type of users, roles, whether in the same institution or different institutions.

**Communication:** Communication between bureaucrats can be done directly between them by tools such as email or phone, or have a centralized system to handle requests or communicate for coordination.

**End-user interaction:** The citizens can interact directly with a system and its output can trigger an outcome, or they are only beneficiary of the outcome while the interaction is made by a system user.

Table 1. Organizational structures in each case.

| Case | VioGen | NAV algorithm | SlimmeCheck |
|---|---|---|---|
| **Managerial** | Centralized in a national unit | Centralized in a national unit | Centralized in a project office |
| **Deployment** | Decentralized in different security bodies | Centralized in a digital unit | Centralized in a project office |
| **Territorialization** | Decentralized in local offices | Remote working | Locally implemented project |
| **System-User Interaction** | Different types of users with access restriction | One type of users | System users do not interact directly with the system |
| **Communication** | Centralized communication in the system | Centralized communication with digital tools | Bureaucrats interact by informal mechanisms |
| **End-User Interaction** | Users do not interact directly | User interacts with the system directly but do not receive system output | User do not interact with the system at all |

## 6. Discussing typologies

During the focus group, the contents of the typologies were discussed. We summarize the main debates and testimonies from participants classified in each of the dimensions.

### 6.1. Capacities

Since most of the systems are based on statistical methods, there is a difference between the case-by-case approach done manually by bureaucrats and the ADMS outcome. Bureaucrats should be aware and should understand how these systems work to be able to consider and critically assess the result. For example, in a risk assessment instrument like VioGen, understanding how a probability is calculated might affect the interpretation of a risk level (Elwood 2016; Heilbrun *et al.* 1999). In spite of these findings, interviewed people were not trained in statistics and it was not considered by managers since they "should be following instructions, not interpreting the risk level" (participant JSR), or in the AMS case " it's not their kind of expertise to know exactly how such a model works" (participant JS). From the provided information that can be offered by the system, it can be assumed that bureaucrats should to interpret and understand the output. Although, model explainability techniques might not be sufficient under a legal aspect (Selbst and Barocas 2018) and social aspect (Miller 2019).

If there are differences from the system outcome and their own assessment without a clear explanation, it could reduce the confidence towards the system. "So, it's kind of hard for us because we automatically want to go closer in and look for problems or challenges that have come up, because we can give some help. But with limited resources it's important to know and trust the statistics", clarified ML.

All the interviewed participants have been working longer before than the implementation of the system, meaning that they already knew how decisions were made before the automation. At the time of implementation, both systems were deployed with almost no training. VioGen included a manual, and over time, they developed a training course. Similarly, the AMS included a manual and a set of explanatory meetings with the involved staff. Instead, the NAV system still does not include any developed training program at national level.

By organizational reasons, not all bureaucrats have the same skills and capacities. For example, in both nation-wide cases training was not distributed evenly in the territory. In the NAV case, only some territorial managers offer a brief training to new bureaucrats. While the VioGen training works as a cascade model, where the central administration trains provincial trainers that provide capacitation course to street-level bureaucrats. In both cases, depends on the territorial authorities to give enough time and resources to staff to be trained. In this regard, VioGen team has been improving how agents are trained and they keep gaining capacities and attending to different courses along their careers. Another fact is that for those that is not mandatory using the system, are less prone to dedicate time to learn and see the potential of it. This is not isolated to only this particular situation because the heavy workload that is common across administrations, also affects the perception of the ADMS (Chien *et al.* 2018).

Lastly, some authors noticed that factors like personal motivations and background influence the adoption of technologies in the working routine (Althoff *et al.* 2017; Kulju *et al.* 2019). While some were comfortable working from home to "avoid friction in social interactions" (participant GA), others claimed that they do not use their "best skills" anymore.

**6.2. Organization**

As pointed in Neumann et al. (2024) setting up an ADMS and AI systems requires to dedicate specific resources. In both cases, the development, research and improvement efforts of the systems are done in collaboration between the office in charge and third parties. This requires not only cooperation but the maintenance and continuous improvement of the system with a permanent allocation of specialized resources. In both cases, results in a centralization of decisions and establishing a team dedicated to the ADMS. At the same time, this explains that updating and improving the system is slow and complex, taking several months or years to release major improvements. Thus, once a system is implemented and users are trained, it is unlikely big changes are made.

The centralization of tasks into a unit that often takes to implement an ADMS does not imply that updating the system and gathering feedback from bureaucrats becomes easier. On the contrary, Participant GA argued, "in the central unit, they are self-fragmented (into different departments). So, some groups work with the algorithm on one thing, and another group is working on another algorithm based on something else, and they don't interact". This causes that information becomes siloed. In addition, centralization implies that a direct contact with territorial teams is lost. Thus, it was suggested to have also different structures where development is spread along the territory and offices.

In the same sense, large systems are not desirable. Creating systems to cover all the other systems could make the administration to collapse. Instead, it was recommended to start from small algorithmic systems that can be implemented on existing systems and scale them up over iterations. Hierarchies and roles are defined in the organization, whether accessing different information, or to oversee a general situation. This is important and affects the perception of administrative staff. Usually street-level bureaucrats have less information about the capabilities of the systems, probably because higher hierarchies consider that is not important. Nevertheless, in both cases bureaucrats were listened in the processes of improvement, participating in research activities and providing feedback when needed.

Beyond maintenance, systems require continuous monitoring when decision making has a critical impact on people's life. For example, VioGen risk evaluations for those cases with high or extreme risk are individually assessed. Similarly, in the NAV case for those cases that are considered very critical and need further assistance a new category was created to attend special needs. For the AMS, due to the low number of daily cases, a brief manual check has be done to most of the cases during the pilot period.

### 6.3. Implementation

During the development and implementation phases of the system, the improvement should not only be focused on the ADMS output but also on systematizing and standardizing input data. This process of standardization also implies normalizing the service across the country and connect different types of professionals and administrations that were not in contact so often (e.g. a police risk evaluation is complemented with a forensic evaluation and both type of professionals is connected by the system in a single interface). In the AMS case, the implementation was hard because there was no standard protocol and generated several internal frictions. However, the process itself helped to reduce flaws in the administrative procedure.

The use of such a system has also weak points. In both cases, it was mentioned that the input might be biased. A reiterated cause was that the citizen might be unable to answer correctly. There are multiple causes for this, depending on the case and more research might be needed to understand in what situations a bureaucrat uses to verify or override the decision. For example, citizens often do not ask for help but solutions to their current situation (e.g. to protect a gender violence victim from an aggressor, or assistance finding a job). But systems are made to evaluate them in order to facilitate the bureaucrats' work, not to offer a solution. Then, the questions made to the citizens are not always matching for what citizens expect from the administration and resulting in mislead or biased answers. Consequently, systems were updated to avoid unexpected inputs from citizens, reducing erroneous or biased outputs. As TG expressed, "If you don't get to talk to people, then you never know. And you get these misplace things, people are misplaced, and we have to do it manually".

In addition, this could lead to biases in the system and alter the final decision. First, because since all human interaction is biased towards a particular interpretation of facts (Tolan 2018). Second, because the decision-making process is expected to discretionally separates one group for another (Calders and Žliobaitė 2013). These ambiguities and interpretation issues plays a role in how people game the system altering their input, expecting an specific outcome from it (Portela *et al.* 2024). This led to the issue that some bureaucrats does not trust in citizens' data because usually citizens are looking for help but the system ask certain questions that are codified and provides a specific output that could not be what the citizen is expecting. Therefore, as TG expressed, "we are other those who should find those who need more help".

In the same sense, bureaucrats can game the system to obtain specific outcomes because they want to help citizens and foresee that the system is not providing the right answer. This finding was also discussed during interviews and in previous research (Portela *et al.* 2024).

There is always a trade-off in the information they gather and use to help citizens. Sometimes citizen input is not helpful, and complementary data can be used to clarify their situation. However, sensitive data such as health data is not allowed by GDPR, and the user's input is commonly ambiguous or biased. NAV is specifically worried on how to solve this situation.

Lastly, ADMS classification categories are only unveiled to the citizens if the administration believes it can help them. For example, informing the risk level of VioGen could help for providing auto-protection measures for violence victims. Rather, in the NAV case, category is not communicated to the citizen because they will ask for more help than expected. In both cases, each category and classification level is always accompanied with the provision of specific resources. Regardless the precision of the algorithm the amount of resources dedicated to give a response should be adequate. For example, the number of resources dedicated for protecting a victim of gender violence should never be lower than what is defined in by the law. Because protect the victim is their duty mandated by law, the number of dedicated staff has been increased year-over-year as a response to the demand increase.

### 6.4. Confidence

When bureaucrats are not involved in the implementation phase and are not considered on the improvements their confidence towards the system and motivation to oversee the ADMS could be reduced (Chien *et al.* 2018; Noori 2021). This was not the case in our interviews. All the participants agreed on a high level of confidence towards the system. No one had a vision of a perfect system because there were some errors and misclassification. But, the perception was mainly related to the effort made to continually improve the ADMS and the complementary tools used together (e.g. databases, search engines, messaging, and alert systems). Managers were expected to have a higher confidence on the system because they might be proud about the result of their work. However, they also were transparent about the need of improvements in the systems.

Much of the perception is altered by the bureaucrat's expectation about the system's outcome. As GA said during the focus group, "if we think that this will, this will work from day one, and this will make everything easier, then we will be really surprised to find that it doesn't sometimes […] if I'm not satisfied with the today's ADMS enough, is because I expected otherwise, I expected something else."

In the VioGen and NAV cases, it was acknowledged that some bureaucrats with less knowledge about how system works tend to think that systems are smarter than what really are, even when the system is only based on rules. On the contrary, the AMS investigators distrust the algorithm output because the explanatory features were not sufficient or not described as they expected, contrasting with their subjective assessment. Nevertheless, we could not collect much evidence about overreliance or under reliance regarding the systems' outputs.

Along the focus group, the topic about bureaucrat's expectation was also related to the skills and capacities to understand the system inner works. It could be thought that as more someone knows how the system works, more confidence can have towards it, moderating their expectations. However, it can be the opposite if they are disappointed from the limited capacities, the amount of workload that it brings or the slow changes during to make it work as expected for solving a specific task. "So, it's not a bullet proof system. They need to know how it works and how it doesn't work" as ML pointed out.

### 6.5. Use

All these technological tools are expected to alleviate their workload. In AMS the workload changed, reducing the work for some bureaucrats and increasing for others, but without a critical impact. In the NAV case, the demand usually increases during the first days of the week, and the negative labour situation can overload the capacities of the digital teams. In VioGen, the system requires to fill long forms with the victim in the police station, making it a long process that could be avoided. Even if the system is there to alleviate their work, some bureaucrats do more than what is expected because they want to help the citizens. Bureaucrats use to collaborate when facing these issues, for example, by splitting their work to be more efficient, to face the increase of demand or to systematize the job.

The workload caused by the system is relative. As LT pointed out "at the end the balance is positive, but if you look at specifics, I don't know specific people specific department, this is different ". In a different aspect, workload may be temporarily increased by unadjusted ADMS, in a transition phase that creates new problematic or unexpected situations. Nevertheless, usually automation reduces the former manual tasks, and coordinated with communication tools usually facilitate the work from bureaucrats. Systems can complement informative or corrective measures to prevent known errors from humans or the algorithms to reduce bureaucrats' workload burden.

How systems are designed and implemented differ from how are used. As GA expressed "I do not believe that the best way that the fully finished thing is finished at the bureaucracies table […] We withdraw it if we need or change it, if we need, we can't, we can't rely upon only brains, we must see how it works. We have to see how people react." It is important to test and refine the system in-the-wild (Crabtree and Chamberlain 2013; Rogers and Marshall 2017). However, as TG expressed "So if you test it intensively, that's a good thing to account. Just have to end the testing sometimes and put it into use. And then you can change more after that". For example, in NAV, they know that some answers are not always well answered, so after testing it, they knew how to change the question or to have an alternative procedure to deal with the wrong outcomes. As ML said, "But as long as we know the weaknesses it's not a problem, because we always do have a look at each case".

They might complement their work with different other systems such as internal messaging systems, or external services as online translators. Messaging or digital meetings are tools that help not only to coordinate activities but also to socialize, especially for those doing remote work. However, some interviewees agreed that face-to-face meetings with citizens still are important, especially for those that are in a critical situation or do not have skills to overcome their situation (e.g. youth jobseekers). Communication is not always desired, as in the AMS, bureaucrats increase their interaction because of the lack of clarity in the decision.

## 7. Conclusions of this work

In this research, we covered different dimensions to understand how the nature of these connections are. We found that long term ADMS become adopted when bureaucrats understand the benefit, as well collaborate to improve it. We also learnt that ADMS becomes an opportunity to increase cooperation and communication between administrations and to standardise processes. Most of the findings in the interviews were confirmed during the focus group. Even with the limitations in the number and representativeness of participants we expanded some of the topics identified in our analysis. It becomes evident the importance of relationships between dimensions.

Answering RQ1, we highlight the importance of developing training for street level bureaucrats that are accompanied with resources and directives that make it possible to reach all the necessary staff. As discussed, training should go beyond software oriented, but also explain causes, values, and reasoning of how the system works. This could not only provide more critical skills to bureaucrats but also provide a consistent narrative, influence a positive attitude towards the system as suggested by Kloft et al. (2024).

About RQ2, we acknowledged technological, social, and organizational arrangements in two different countries and contexts of use. Despite the fact that our conclusions could be limited number of cases, we found similarities in both cases regarding their organization (e.g. both systems are centralized, much of the implementation depends on territorial managers), technologies (the system also permit communicate between bureaucrats along the territory, classification evolved over time increasing the categories), social dynamics (face-to-face encounters with the citizen is key in critical scenarios) and legal arrangements (both are solving legal mandates). Nevertheless, we also found several differences, as the different type of ADMS act differently in their output, or the fact that the output is shown to the citizen in one case and not in the other.

Both addressed research questions constitute important elements of the infrastructuring of ADMS in the public sector. In our studied cases, both training and arrangements are parts of this process that generate future-proof systems. As le Dantec and Disalvo expressed "Infrastructuring enables a public's members to identify and address issues in an ongoing manner, creating a socio-technical response that relates the current context of the public to the future context the public is trying to attain"(le Dantec and DiSalvo 2013, 258). At the same time, the different roles are key to understand that their vision could differ in terms of implementation and organizational schemes. As Pipek and Volker pointed, the contribution of different members is important form of infrastructuring, "understood as reconceptualizing one's own work in the context of existing, potential, or envisioned IT tools, is a natural part of every user's activities"(Pipek and Volker 2009, 469).

## 8. Limitations

Despite these interesting findings, we must acknowledge several limitations on our research. We acknowledge the few cases involved in the research; thus, results are difficult to generalize to other specific ADMS and could oversee many details and internal processes on the studied ADMS. Due to the lack of time, we also acknowledge the limitation to study deeply some of the factors identified during the interviews. It resulted very difficult to find use cases that still are in place and street-level bureaucrats for interviewing. For different reasons many systems were cancelled, and we got few responses from administrations. Despite being a pilot, we included the AMS "Slimme Check" case because we consider that it is a good example and we found that bureaucrats received the same training than other cases, and because it was a system developed during almost 5 year. At the time of this article being written, the Slimme Check project was formally cancelled and the team no longer work at the administration. The decision was made based on the lack of clarity on how biases might affect citizens. Due to organizational circumstances, participants from SlimmeCheck and VioGen did not participate. This unfortunate result was despite the efforts made to invite them to participate, exceeding our capacities to engage them. From the proposed methodologies, the artifact mapping and analysis was not fully developed because interviews took more time than expected. Therefore, a deep exploration can be done in future research. In addition, we couldn't have the opportunity to discuss their perception on automation but we observed a positive view about accepting a limited automation with a grade of autonomy for bureaucrats, that was mentioned in previous studies (Cummings 2006; Portela *et al.* 2024).

**Acknowledgments.** The authors are grateful for the continuous support and collaboration provided by the members of the Amsterdam City Council, the Interior Ministry from Spain and the Norwegian Labour and Welfare Administration (NAV). We also thank all the other administrations that also contributed to the first part of this research. In addition, Manuel recognizes for their continuous support to Antonio Andres-Pueyo from Universidad de Barcelona, Carlos Castillo (Chato) and Vladimir Estivill-Castro from UPF."
**Funding statement.** This work has been partially supported by the Public Sector Tech Watch programme, Joint Research Centre, European Commission through the Expect contract CT-EX2019D347180-103.
**Competing interests.** None.
**Data availability statement.** None.
**Author contributions.** Conceptualization: M.P; L.T; P.R.M Methodology: M.P; L.T; P.R.M. Data curation: M.P; L.T; P.R.M. Formal Analysis: M.P. Data visualisation: M.P. Writing original draft: M.P. Supervision: L.T; P.R.M. All authors approved the final submitted draft.